\documentclass[aps,twocolumn,groupedaddress,superscriptaddress,floatfix,showpacs,amsmath,amssymb,prb]{revtex4}
\usepackage{amsmath}
\usepackage{amssymb}
\usepackage{graphicx}
\usepackage{bm}	

\usepackage[usenames,dvipsnames]{color}	

\begin{document}
    %\linenumbers
    \title{ Long-distance radiative coupling between quantum dots in photonic crystal dimers }

\author{J. P. Vasco}\email{jpvasco@gmail.com}
\affiliation{Departamento de F\'{\i}sica, Universidade Federal de Minas Gerais, Belo Horizonte MG, Brazil}
\affiliation{DISSE - INCT de Nanodispositivos Semicondutores, Brazil} 
\author{P. S. S. Guimar\~{a}es}\email{pssg@fisica.ufmg.br}
\affiliation{Departamento de F\'{\i}sica, Universidade Federal de Minas Gerais, Belo Horizonte MG, Brazil}
\affiliation{DISSE - INCT de Nanodispositivos Semicondutores, Brazil} 
\author{D. Gerace}\email{dario.gerace@unipv.it}    
\affiliation{Dipartimento di Fisica, Universit\`a di Pavia, via Bassi 6, I-27100 Pavia, Italy}

    %\date{\today}
\begin{abstract}
We study the mutual interaction between two identical quantum dots coupled to the normal modes of two-site photonic crystal molecules in a planar waveguide geometry, i.e. \textit{photonic crystal dimers}. We find that the radiative coupling between the two quantum emitters is maximized when they are in resonance with either the bonding or the antibonding modes of the coupled cavity system. Moreover, we find that such effective interdot coupling is sizable, in the range of $\sim 1$~meV, and almost independent from the cavities distance, as long as a normal mode splitting exceeding the radiative linewidth can be established (strong cavity-cavity coupling condition). In realistic and high quality factor photonic crystal cavity devices, such distance can largely exceed the emission wavelength, which is promising for long distance entanglement generation between two qubits in an integrated nanophotonic platform. We show that these results are robust against position disorder of the two quantum emitters within their respective cavities.
\end{abstract}

\maketitle

\section{Introduction}
There is growing interest in controlling the radiative coupling between distant quantum emitters in integrated photonic technologies \cite{qp_review}, with the main aim of realizing two-qubit gates in a reliable architecture for quantum information processing \cite{imamoglu99}. Semiconductor quantum dots (QDs) currently represent very promising candidates to implement single and two-qubits operations, owing to their large oscillator strength and long decoherence time scales~\cite{warburton_review}. 

In general, the mutual interaction between two QDs decays rapidly when their distance is larger than the emission wavelength \cite{parascandolo2005}. Possible ways to enhance the radiative coupling between two distant QDs have been proposed in the last few years, e.g. exploiting electromagnetic field confinement in planar microcavities \cite{tarel2008} or photonic crystal (PC) integrated circuits \cite{hughes2007prl}. In fact, PCs in planar waveguides, or PC slabs, have emerged as the preferential nanophotonic platform in view of realizing a fully integrated quantum photonic technology, owing to their engineering flexibility to tailor the propagation and confinement of light at optical or near-infrared wavelengths \cite{akahane2003,notomi2004,notomi_review}. The strong light-matter coupling regime between a single QD and a cavity mode requires large oscillator strength of the dipole emitter, and a small cavity mode volume \cite{andreani99prb}. 
Such characteristics can naturally be fulfilled in PC cavities \cite{yoshie2004}, where QDs with large dipole moment can be deterministically positioned at the field antinodes \cite{badolato2005}; the quantum nature of such a system has now been fully established \cite{kevin07nat,faraon08nphys,reinhard2012nphot}. In addition, PC cavities have already been assessed for broadband and fine post-processing tuning \cite{hennessy2005}, and tested against intrinsic disorder \cite{gerace05pnfa,portalupi2011}. Recent advances in cavity design have led to remarkable figures of merit, such as flexible far-field engineering \cite{tran2009,portalupi2010opex}, as well as genetic optimization of ultra-high Q-factors \cite{minkov2014scirep}, with experimental values on the order of 2 million \cite{laiAPL2014}. 

Proposals for increasing the mutual interaction distance between two QDs in a PC platform mainly considered using a waveguide as a bus for photon propagation \cite{hughes2007prl,yao2009,minkov2013a}. The role of disorder on light localization was also addressed \cite{minkov2013b}. Alternatively, preliminary studies considered the mutual coupling between two QDs positioned at the field antinodes within the same PC cavity \cite{imamoglu2007,minkov2013a}, for which early experimental evidence was shown \cite{postigo2010}. The possibility of mediating the inter-QD coupling through the normal modes of a photonic molecule has been considered for coupled micro disks \cite{armando2008apl}, where the distance is limited by evanescent inter-cavity coupling in free space. 

\begin{figure}[b]
  \begin{center}
    \includegraphics[width=0.45\textwidth]{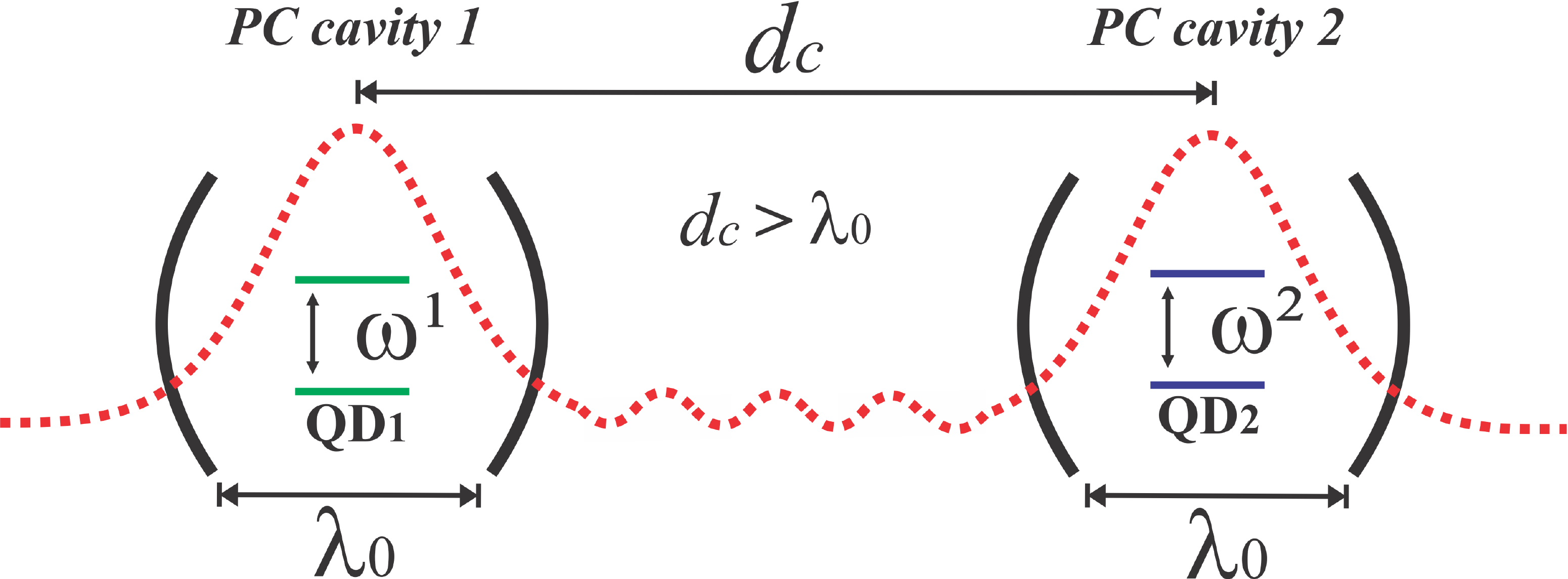}
  \end{center}
  \caption{(Color online) Schematic representation of the system investigated in this work: two strongly coupled nanocavities, each containing a single QD. The distance between the nanocavities, $d_c$, can be larger than the characteristic QD emission wavelength in vacuum, $\lambda_0$.}
  \label{fig:Scheme}
\end{figure}

Here we theoretically address the possibility of using strongly coupled PC molecules to efficiently increase the mutual QDs coupling rate even at large distance, which has been overlooked in the literature, so far. A schematic representation of such a system is shown in Fig. \ref{fig:Scheme}. The photonic molecules we are interested in this work are composed of two coupled PC slab cavities \cite{ishii2006}, or PC dimers. We treat the light-matter coupling in a semiclassical formalism based on Green's tensors, following the formalism developed in Ref.~\onlinecite{minkov2013a} for generic PC structures. The classical electromagnetic fields are solved within a guided-mode expansion approach \cite{andreani06prb}. Interestingly, PC dimers present peculiar characteristics, such as coupling strength increasing with distance and switching of the fundamental mode bonding/antibonding character \cite{kee2003prb,caselli2012}. In particular, we will specifically consider coupled L3 cavities, \textit{i.e.}, three missing holes in a 
hexagonal lattice \cite{akahane2003}. The coupling characteristics of the normal modes in such structures have already been addressed experimentally \cite{atlasov2008,chalcraftOpex}. A high degree of control on mode tuning has now been demonstrated with a variety of fabrication techniques \cite{kicken2009opex,caselli2012apl,intonti2012apl,waks2013apl}, and the possibility of reversing the mode symmetry has been shown \cite{caselli2014opex,yacomotti2014}, which places these systems among the best possible candidates to practically explore the rich physics of coupled quantum modes.

The advantage of using of PC dimers to mediate the mutual QDs coupling is twofold. First, both bonding and antibonding mode profiles are strongly confined in the two cavities, which allows to maximize the QD-normal mode coupling for each emitter. Second, we find that the two QDs placed in the two different cavities of the PC dimer are radiatively coupled with an effective rate that essentially does not depend on distance, as long as the photon exchange rate between the two cavities is larger than the photon escape rate (or, in other words, as long as the normal mode splitting between bonding and antibonding modes can be spectrally resolved). In PC dimers, we show that this distance can be significantly larger than the wavelength of the fundamental cavity mode. Moreover, given the recent demonstration of ultra-long-distance inter-cavity coupling in a PC chip \cite{noda_nphot2012}, the conclusions above unequivocally solve the issue of mutual radiative coupling of distant QDs for application in quantum 
information processing.  
We also notice that the present results can be of interest in view of applying coupled cavity systems to explore the rich interplay between coherent inter-cavity tunneling and quantum nonlinearities, as proposed in a few recent works \cite{gerace_josephson,ferretti2010,savona10prl,bamba}. The strong coupling of a single QD coupled to a PC dimer has already been addressed experimentally \cite{arka2012prb}. More directly connected with the present study, a recent proposal envisions the possibility of achieving a steady state entanglement between distant qubits, just exploiting the selective pumping of the two cavities in a photonic dimer \cite{hakan_arxiv}. Implementing the latter proposal with two quantum dots mutually coupled through a PC dimer can be crucial to optically address each cavity separately from the other, since the inter-cavity separation can be larger than the resonance wavelength in such systems.

The paper is organized as follows. In Sec. \ref{model}, we present the key aspects of the semiclassical method formulated in Ref.~\onlinecite{minkov2013a} for arbitrary PC structures. The PC dimers in which we are interested are characterized by using the Guided Mode Expansion (GME) approach \cite{andreani06prb} in Sec. \ref{dimer}. In Sec. \ref{QDPCdimer}, we study the polariton states and the radiative coupling between two QDs coupled to the PC dimers, and we use a statistical analysis to study the effects of the non-ideal positioning of the two QDs. Finally, in Sec. \ref{conclusions} the main conclusions of this work are presented. 

\section{Theoretical model}\label{model}
A useful semiclassical formalism was described by Minkov and Savona in Ref. \onlinecite{minkov2013a} to study $N$ QDs coupled to $M$ electromagnetic photonic modes in an arbitrary dielectric structure. Following this work, our starting point is the inhomogeneous wave equation for the electric field, with a polarization vector source that takes into account the linear optical response of the QDs (low excitation regime) through a nonlocal susceptibility tensor. The electromagnetic problem is solved through the Green's function approach. The field eigenmodes are used to expand the Green's tensor in the resolvent representation, and the wave function $\Psi_{\alpha}(\mathbf{r})$ of the QD $\alpha$ is represented using a point dipole assumption, which is a good approximation when the electric field does not vary significantly in the region where $\Psi_{\alpha}(\mathbf{r})$ is non-negligible. Hence, the set of equations that define the complex frequency poles of the coupled QD-photonic system (polariton frequencies) can be given as \cite{minkov2013a}:
\begin{equation}\label{syseq}
(\omega^{\beta}-\omega)\widetilde{\mathbf{Q}}^{\beta}(\omega)=\sum_{\alpha=1}^{N}\sum_{m=1}^{M}\frac{\mathbf{g}_{m}^{\beta}\otimes\mathbf{g}^{\alpha *}_{m}}{(\omega_m-\omega)}\widetilde{\mathbf{Q}}^{\alpha}(\omega),
\end{equation}
where $\widetilde{\mathbf{Q}}^{\alpha}(\omega)=\mathbf{Q}^{\alpha}(\omega)/(\omega^\alpha-\omega)$, and $\mathbf{Q}^{\alpha}=\int d\mathbf{r}\Psi_{\alpha}(\mathbf{r})\mathbf{Q}(\mathbf{r},\omega)$ are the overlap integrals between the wavefunction of the QD $\alpha$ and the field $\mathbf{Q}(\mathbf{r},\omega)$, with $\mathbf{Q}(\mathbf{r},\omega)=\sqrt{\epsilon(\mathbf{r})}\mathbf{E}(\mathbf{r},\omega)$. In the present formalism, the transition frequencies of the QDs are denoted by superscripts, while the photonic eigenmodes frequencies are denoted by subscripts. The outer product $\otimes$ is defined as:
\begin{equation}\label{outer}
 \mathbf{A}\otimes\mathbf{B}=
 \begin{pmatrix}
  A_xB_x & A_xB_y\cr
  A_yB_x & A_yB_y
 \end{pmatrix} \, .
\end{equation}
The quantities $\mathbf{g}^{\alpha}_{m}$ in Eq.~(\ref{syseq}) are the coupling strengths between the $m$th photonic mode and the $\alpha$th QD, and they are defined as follows:
\begin{equation}\label{gsdirac}
\mathbf{g}^{\alpha}_{m}=\left(g^{\alpha}_{m,x},g^{\alpha}_{m,y}\right)=\left(\frac{2\pi \omega_0}{\epsilon_\infty\hbar}\right)^{1/2}d ~\mathbf{Q}_m(\mathbf{r}_{\alpha}),
\end{equation}
where $\omega_0$ is an average exciton transition frequency, $d$ is the dipole moment of the QD, $\epsilon_\infty$ is the dielectric constant of the semiconductor and the $m$th mode is evaluated at $\mathbf{r}_{\alpha}$, the position of the $\alpha$th QD. To obtain Eq.~(\ref{gsdirac}), we have used explicitly the point dipole assumption in the QD wave function, \textit{i.e.}, $\Psi_{\alpha}(\mathbf{r})=C\delta(\mathbf{r}-\mathbf{r}_{\alpha})$, with $C$ a normalization constant. From Eq.~(\ref{syseq}) we define the following tensor:\\

\begin{equation}\label{Gtensor}
\hat{G}^{\alpha\beta}(\omega)=\sum_{m=1}^{M}\frac{\mathbf{g}_{m}^{\beta}\otimes\mathbf{g}^{\alpha *}_{m}}{(\omega_m-\omega)}=d^2\frac{2\pi}{\epsilon_\infty\hbar}\frac{\omega^2}{c^2}\hat{G}(\mathbf{r}_{\alpha},\mathbf{r}_{\beta},\omega),
\end{equation}
where $\hat{G}(\mathbf{r}_{\alpha},\mathbf{r}_{\beta},\omega)$ is the Green's tensor evaluated at the QDs positions $\mathbf{r}_\alpha$ and $\mathbf{r}_\beta$. The components $G^{\alpha\beta}_{xx}$, $G^{\alpha\beta}_{xy}$, $G^{\alpha\beta}_{yx}$ and $G^{\alpha\beta}_{yy}$ are interpreted as the effective radiative coupling strengths between the QD $\alpha$ and QD $\beta$ at the excitonic transition frequency $\omega$. The complex frequency poles of the system are found by imposing the singular condition of the associated matrix of Eq.~(\ref{syseq}), which is mathematically equivalent to diagonalize the matrix \cite{minkov2013a}:
\begin{equation}\label{effmatrix}
\Lambda=
 \begin{pmatrix}
 \omega^1_x & 0 & \cdots & 0 & g^1_{1,x} & \cdots & g^1_{M,x}\cr
 0 & \omega^1_y & \cdots & 0 & g^1_{1,y} & \cdots & g^1_{M,y}\cr
 \vdots & \cdots & \ddots & \vdots & \vdots & \cdots & \vdots \cr
 0 & 0 & \cdots & \omega^N_y &  g^N_{1,y} & \cdots & g^N_{M,y} \cr
 g^{1*}_{1,x} & g^{1*}_{1,y} & \cdots & g^{N*}_{1,y} & \omega_{1} & \cdots & 0\cr
  \vdots & \cdots & \ddots & \vdots & \vdots & \cdots & \vdots \cr
  g^{1*}_{M,x} & g^{1*}_{M,y} & \cdots & g^{N*}_{M,y} & 0 & \cdots & \omega_{M}
 \end{pmatrix},
\end{equation}
where possible deviations from the perfectly symmetrical QDs can be introduced in the model through different transition frequencies in the $x$ and $y$ directions, respectively. In this way, the real part of the eigenvalues of $\Lambda$ are the polariton frequencies of the coupled system, and the imaginary part determines their loss rates. The corresponding eigenvectors define the Hopfield coefficients \cite{hopfield1958}, i.e. $\bm{\lambda}=\left(\lambda^1_x,\lambda^1_y,\ldots,\lambda^N_x,\lambda^N_y,\lambda_1,\ldots,\lambda_M\right)$, whose square moduli can be directly interpreted as the bare-exciton (or bare-photon) fractions of the mixed polariton state.

For typical self-organized InGaAs QDs with exciton transition energy in the $\sim1.3$~eV range, the square dipole moment can be estimated to be $d^2\approx0.51$~eV~nm$^3$, using the measured radiative decay rate of QDs with a radiative life time of 1~ns \cite{minkov2013a}. We adopt such value throughout this work. The last requirement of the model are the values of the normalized photonic eigenmodes $\mathbf{Q}_m$ at the QD positions, and their corresponding eigenfrequencies and losses, which are computed using standard methods to solve PC structures. In this work we employ the GME approach \cite{andreani06prb}, which is the best compromise between computational effort and reliable results for strong localized modes in high dielectric regions, as it is the case of PC dimers.

\begin{figure*}[htp!]
  \begin{center}
    \includegraphics[width=0.8\textwidth]{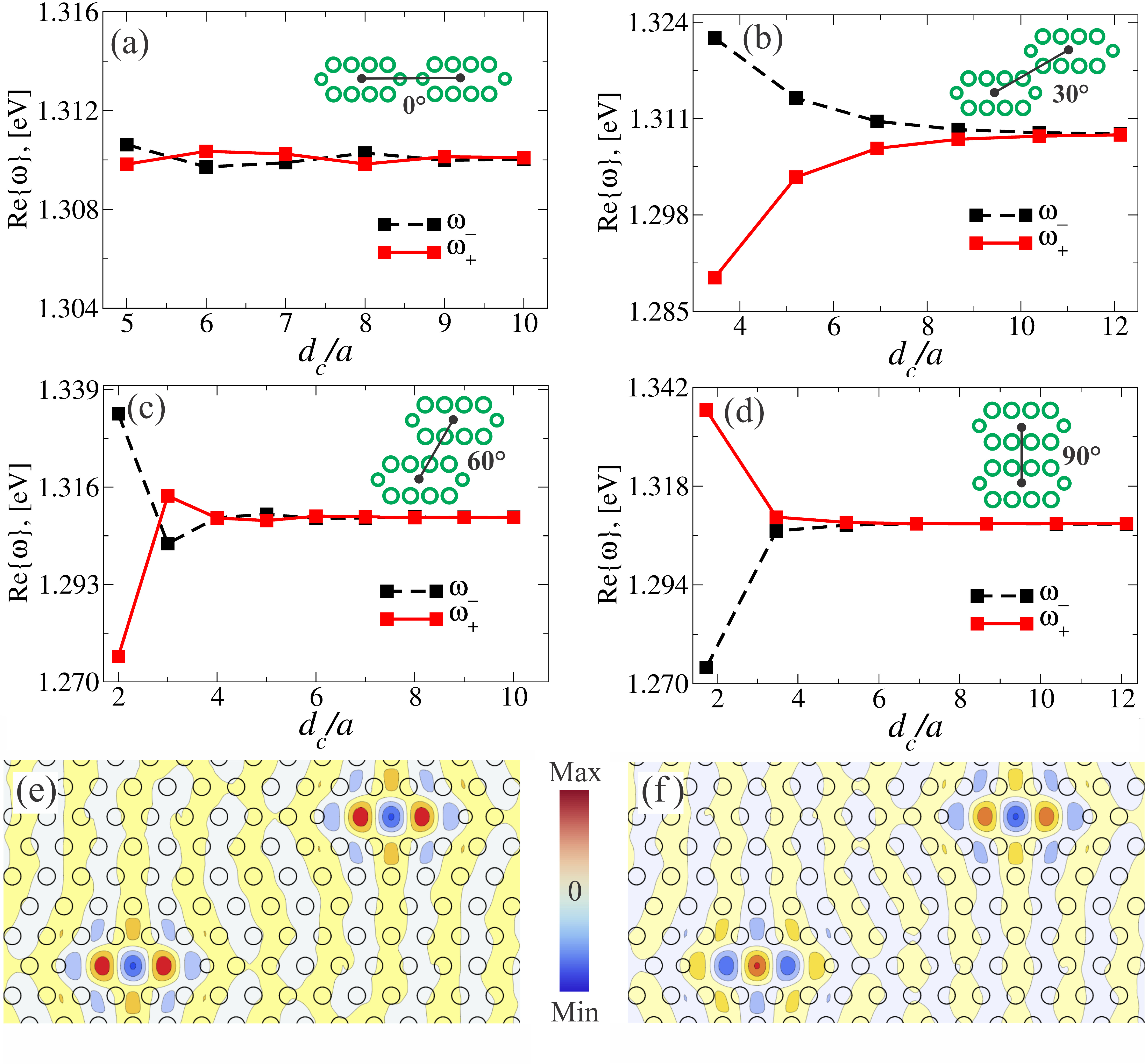}
  \end{center}
  \caption{(Color online) Photonic dispersion of the bonding (solid lines) and antibonding (dashed lines) modes for the {(a) $0^{\circ}$, (b) $30^{\circ}$, (c) $60^{\circ}$, and (d) $90^{\circ}$ cases}, respectively. Electric field components $E_y$ associated to the (e) bonding and (f) antibonding states for the $30^{\circ}$ PC dimer at $d_c=5\sqrt{3}a$.}
  \label{fig:PC-structure}
\end{figure*}

\section{Photonic crystal dimers}\label{dimer}
The PC dimers that we address in this work are formed by two L3 cavities embedded in a PC slab with a hexagonal lattice of holes. The two cavities are nominally identical and have an outward displacement of the two lateral holes $s=0.15a$ along the cavity axis, where $a$ is the PC lattice constant, and their radii are decreased to 80\% of the surrounding PC holes radius, which improves significantly the quality factor \cite{gerace04pnfa}. The L3 cavities are disposed in such a way that the line connecting their centers makes an angle of {$0^{\circ}$, $30^{\circ}$, $60^{\circ}$, or $90^{\circ}$} with the horizontal axis \cite{chalcraftOpex}, respectively. We chose the same parameters as in Ref.~\onlinecite{minkov2013a} which are relevant in GaAs structures, namely, lattice constant $a=260$~nm, hole radius 65~nm and slab thickness 120~nm, with real part of the refractive index $\sqrt{\epsilon_\infty}=3.41$. Here, the photonic modes for the {$0^{\circ}$, $30^{\circ}$, $60^{\circ}$, and $90^{\circ}$} PC dimers are computed using the GME method \cite{andreani06prb}.
{We use a hexagonal supercell of superlattice parameter $24a$, and up to 11025 total plane waves tested for convergence in the $30^{\circ}$ and $60^{\circ}$ cases; we use rectangular supercells of dimensions $27a\times8\sqrt{3}a$ and $14a\times19\sqrt{3}a$, and up to 11881 and 14641 total plane waves tested for convergence, in the $0^{\circ}$ and $90^{\circ}$ cases}, respectively. Since we are interested in the frequency region below the second-order mode of the homogeneous slab, we consider only one guided mode in the expansion but we have checked that adding a second-order mode does not affect the results appreciably. Finally, the real part of the frequencies are averaged in the first Brillouin zone of the superlattice in order to smooth out finite supercell effects. 

Figure~\ref{fig:PC-structure} shows the results of the GME computations for the first two PC dimer modes, associated to the split states arising from the fundamental L3 cavity mode. The bonding states are labeled with the subscript $+$, while the subscript $-$ is used for the antibonding states. 
{Panels (a), (b), (c), and (d) correspond to the $0^{\circ}$, $30^{\circ}$, $60^{\circ}$ and $90^{\circ}$ PC dimers, respectively. For the $0^{\circ}$ case we see a very small splitting between the normal modes, which does not change appreciably with the inter-cavity distance $d_c$ (defined from center-to-center of the two PC slab cavities). The behavior of the normal mode frequencies is quantitatively different for the other cases, in which a large splitting at small inter-cavity distances can be noticed. Such splitting decreases smoothly for the $30^{\circ}$ dimer, and much more rapidly for the $60^{\circ}$ and $90^{\circ}$ dimers, on increasing $d_c$. Nevertheless, between $d_c=4a$ and $d_c=5a$ for the $60^{\circ}$ case, and between $d_c=7a$ and $d_c=8a$ for the $0^{\circ}$ case the splitting increases, which is a rather counterintuitive behavior and it typically occurs in PC molecules, as already evidenced \cite{caselli2012,chalcraftOpex}. In addition, the bonding ($+$) and antibonding ($-$) behavior of the modes changes as a function of distance, which is another interesting phenomenon already seen in experimental measurements on such systems \cite{caselli2012}. As it is expected, the resonance frequencies of these PC dimers tend to the values of the isolated L3 cavity for large distances.} In Figs.~\ref{fig:PC-structure}(e) and \ref{fig:PC-structure}(f) we show the $E_y$ patterns for the bonding and antibonding states, respectively, for the case of $30^{\circ}$ at the inter-cavity distance $d_c=5\sqrt{3}a$. The symmetry point of these PC molecules is located at the center of the structure. The bonding (antibonding) mode has an even (odd) symmetry with respect to this point, as it can be seen in the figure. {Since the $30^{\circ}$ and $60^{\circ}$ cases represent the most interesting physical behaviors, we will focus on these throughout the following of this work.}

\begin{figure}[htp!]
  \begin{center}
    \includegraphics[width=0.45\textwidth]{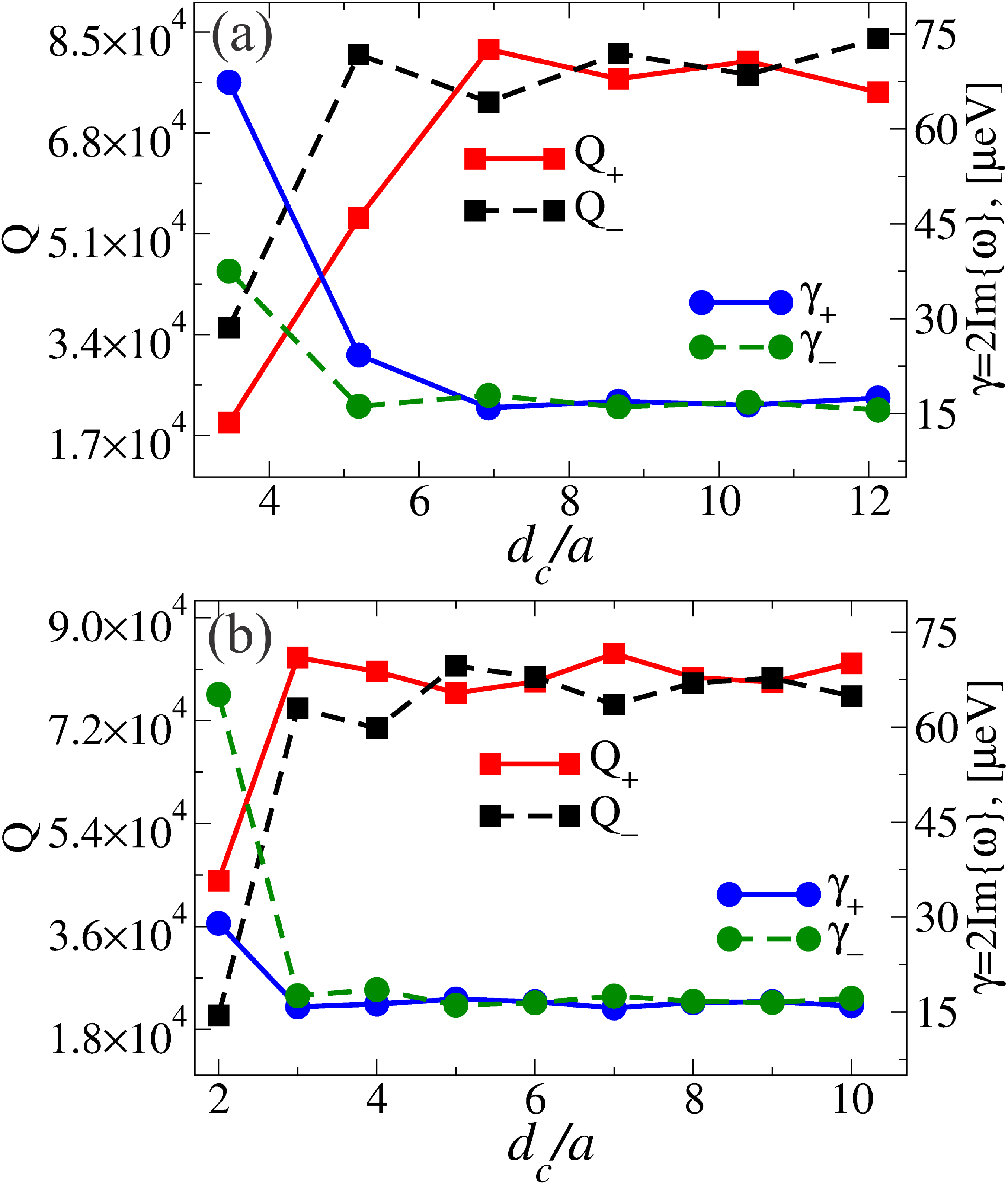}
  \end{center}
  \caption{(Color online) {Quality factors, $Q$ (left axis), and loss rates, $\gamma_m$ (right axis) for the (a) $30^{\circ}$ and (b) $60^{\circ}$ cases as a function of the inter-cavity distance, $d_c$.}}
  \label{fig:Q-factors}
\end{figure}

The imaginary parts of photonic eigenfrequencies are calculated with the photonic Fermi's golden rule using time-dependent perturbation theory \cite{Sakoda}, {and averaging in the first Brillouin zone of the superlattice, in the same way as it was done for the real parts}; the corresponding quality factor is computed with these averaged quantities through $Q=\mbox{Re}\{\omega\}/2\mbox{Im}\{\omega\}$. {Figures~\ref{fig:Q-factors}(a) and \ref{fig:Q-factors}(b) show the quality factors and loss rates, $\gamma_m=2\mbox{Im}\{\omega_m\}$, of the split modes for the $30^{\circ}$ and $60^{\circ}$ PC dimers, respectively. The loss rates are relatively high for large splitting [see Figs.~\ref{fig:PC-structure}(b) and \ref{fig:PC-structure}(c)] and decrease quickly when $d_c$ increases, tending to the values of the isolated L3 cavity; this trend can also be verified in the quality factor, taking into account the inverse dependence with Im$\{\omega\}$. Here, both quality factors and loss rates, tend to the values of the isolated L3 cavity for large distances, as it is expected.} 
The calculations shown in Figs.~\ref{fig:PC-structure} and \ref{fig:Q-factors} agree with previous work on similar systems~\cite{chalcraftOpex}.

\section{Quantum dots coupled to photonic crystal dimers}\label{QDPCdimer}
After characterizing the photonic structures, we have all the parameters required to study the coupled QD-PC dimer system. In this section we study the coupling of two QDs coupled to the PC dimers studied in Sec. \ref{dimer}, using the formalism presented in Sec. \ref{model}.

\begin{figure*}[htp!]
  \begin{center}
    \includegraphics[width=0.8\textwidth]{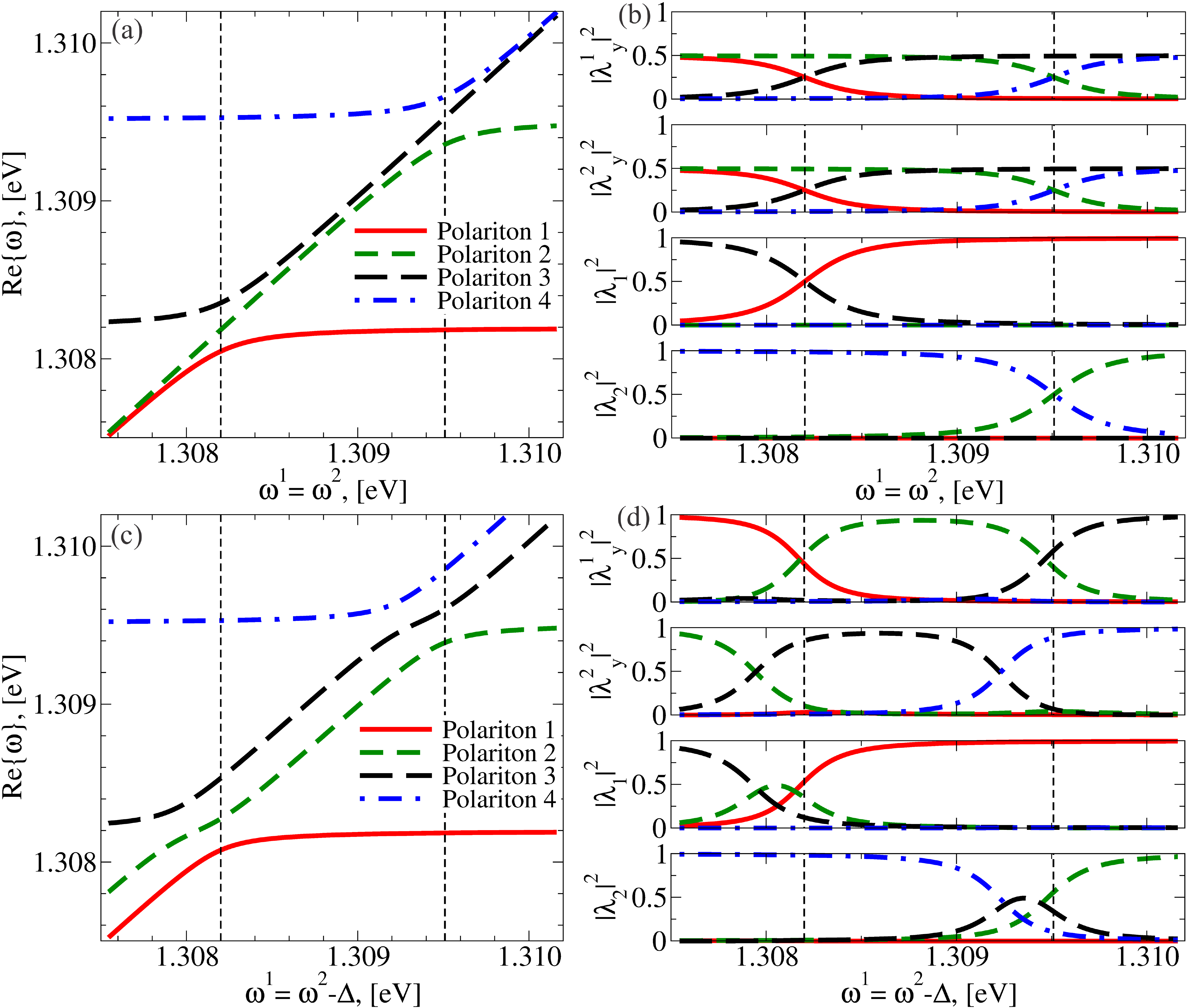}
  \end{center}
  \caption{(Color online) (a) Real part of the eigenfrequencies of the PC-QD system for the $30^{\circ}$ PC dimer at $d_c=5\sqrt{3}a$, for zero dot-dot detuning. (b) The square modulus of the Hopfield coefficients associated to (a). (c) and (d) correspond to the same case as in (a) and (b), respectively, but with a finite dot-dot detuning of $\Delta=300$ $\mu$eV. In panels (b) and (d), $\lambda^1_y$ and $\lambda^2_y$ are associated to the QDs 1 and 2, respectively, and $\lambda_1$ and $\lambda_2$ are associated to the PC modes with frequencies $\omega_1$ and $\omega_2$, respectively.}
  \label{fig:Polaritons}
\end{figure*}

\subsection{Polariton states}
We consider each QD positioned at the center of each L3 cavity, which simplifies considerably the problem since the QDs only couple with the $y$ component of the electric field. This is due to the fact that the $x$ field component is negligible for small inter-cavity distances $d_c$ at the center of each L3 cavity, and eventually tends to zero for large $d_c$ values (the $E_x$ is exactly zero at the center of the isolated L3 cavity). We also consider that the loss rates through the PC dimer modes are significantly larger than the QD loss rates $\gamma^{\alpha}$ through other channels, therefore, we set $\gamma^{\alpha}=0$. With these conditions the $\Lambda$ matrix of Eq.~(\ref{effmatrix}) takes the following simplified form:
\begin{equation}\label{eff4x4}
\Lambda_y=
\begin{pmatrix}
 \omega^1 & 0 & g^1_{1,y} & g^1_{2,y} \cr
 0 & \omega^2 &  g^2_{1,y} & g^2_{2,y} \cr
 g^{1*}_{1,y} & g^{2*}_{1,y} & \omega_1-i\frac{\gamma_1}{2} & 0 \cr
 g^{1*}_{2,y} & g^{2*}_{2,y} & 0 & \omega_2-i\frac{\gamma_2}{2}
\end{pmatrix},
\end{equation}
where $\omega^1=\omega^1_y$, $\omega^2=\omega^2_y$, $\omega_1=\min(\omega_{+},\omega_{-})$ and $\omega_2=\max(\omega_{+},\omega_{-})$. In Fig. \ref{fig:Polaritons}(a) we show the real part of the eigenfrequencies from diagonalization of the matrix in Eq. (\ref{eff4x4}), for the same dimer configuration shown in Figs.~\ref{fig:PC-structure}(e) and \ref{fig:PC-structure}(f), as a function of the frequencies of the two QDs at zero dot-dot detuning. We see that vacuum Rabi splitting occurs in the frequency regions associated to the coupled modes resonances (the vertical dashed lines), which is the signature of the strong coupling between the excitonic QD states and the PC modes. Due to the opposite symmetry of the two photonic modes, i.e. symmetric (bonding) and antisymmetric (antibonding), the relevant QD states in the coupling are the excitonic states with the symmetry of the PC mode. In this way, the first anti-crossing, which is associated to the bonding PC mode, corresponds to an excitonic symmetric state and the 
antisymmetric remains dark, whilst the second anti-crossing, which is associated to the antibonding PC mode, corresponds to an excitonic antisymmetric state with the symmetric one remaining dark. When the first PC mode is antibonding, as in the case of some configurations of the {$0^{\circ}$, $60^{\circ}$ and $90^{\circ}$}  PC dimers, the first anti-crossing corresponds to an antisymmetric excitonic state, and the second one to a symmetric excitonic state, as verified in our calculations. The Hopfield coefficients are shown in Fig. \ref{fig:Polaritons}(b), and we can see an interesting collective behavior suggesting that a strong radiative coupling between the QDs is present at each frequency of the PC dimer modes. In the region of the first strong coupling the polaritons 1 and 3 have comparably significant values of the coefficients $\lambda^1_y$ (QD 1), $\lambda^2_y$ (QD 2), and $\lambda_1$ (mode 1), and we see the same behavior for polaritons 2 and 4 in the region of the second strong 
coupling, but now with the mode coefficient 
$\lambda_2$ (mode 2), which corresponds to the antisymmetric mode. Usually, it is very likely that two QDs are detuned due to their inhomogeneous distribution of sizes.
Therefore, we show in Figs. \ref{fig:Polaritons}(c) and \ref{fig:Polaritons}(d) the same analysis made in panels (a) and (b) but now introducing a finite and sizable detuning between the two QDs, $\Delta=\omega^1  - \omega^2 = 300$ $\mu$eV. We see that under such conditions, symmetric and antisymmetric excitonic states are possible and the dark mode is not present. 
All the Hopfield coefficients associated to the coupling of the QDs with each PC dimer mode are non-negligible, and consequently radiative coupling between the QDs remains present.

\begin{figure*}[htp!]
  \begin{center}
    \includegraphics[width=0.8\textwidth]{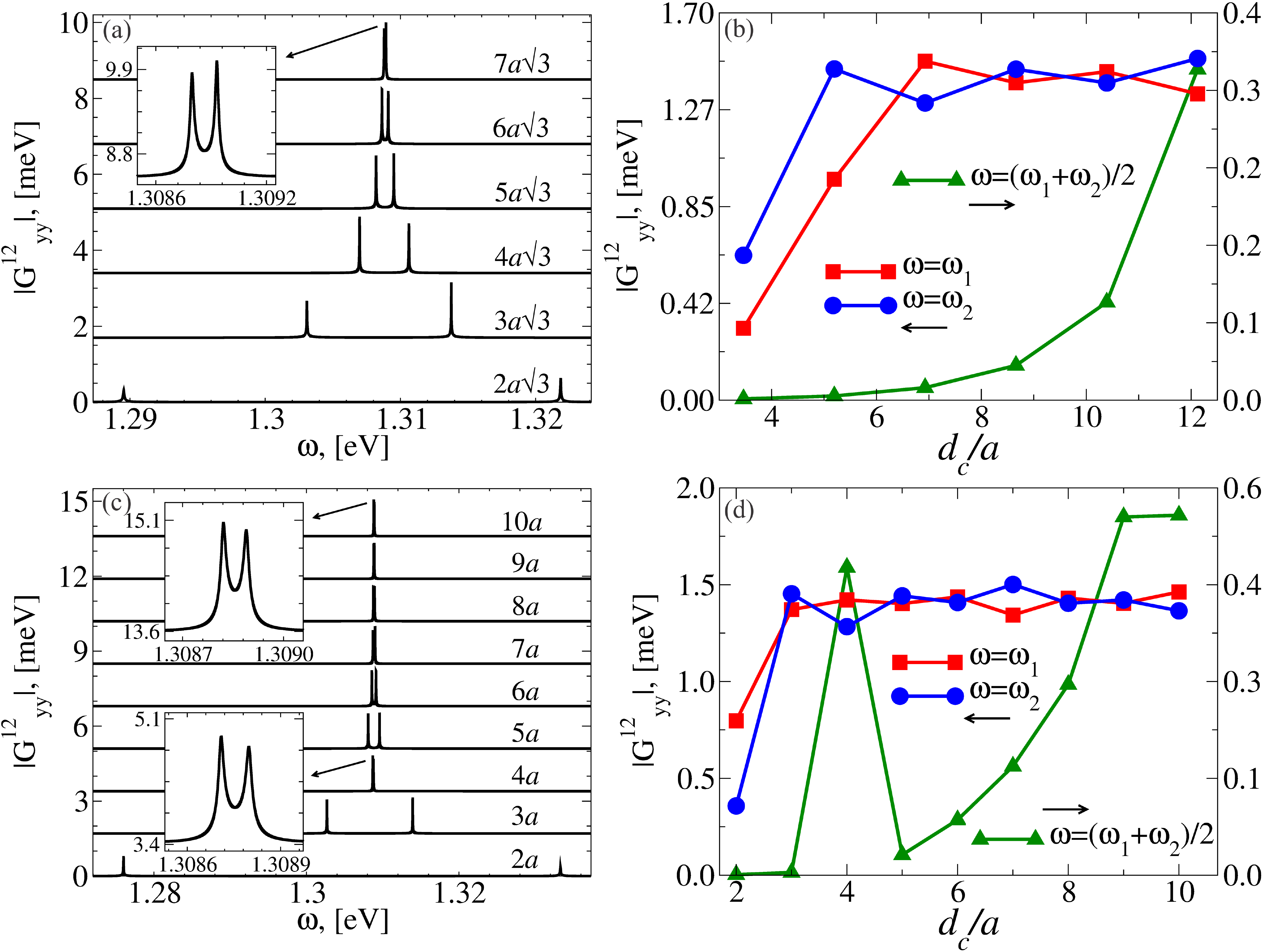}
  \end{center}
  \caption{(Color online) (a) Absolute value of the Green's tensor component $G^{12}_{yy}(\omega)$ for the $30^{\circ}$ dimer as a function of the exciton transition frequency as a function of interdot distance. Each curve is displaced vertically by 1.7~meV from the previous curve; the vertical scale is valid for the smallest interdot distance. (b) Absolute value of the component $G^{12}_{yy}(\omega)$ evaluated at $\omega=\omega_1$, $\omega=\omega_2$ and $\omega=\omega_{av}=(\omega_1+\omega_2)/2$ as a function of the interdot distance. (c) and (d), same as (a) and (b), respectively, but for the $60^{\circ}$ dimer. The insets of panels (a) and (c) evidence the two splitted peaks in $|G^{12}_{yy}(\omega)|$.}
  \label{fig:GreenF}
\end{figure*}

\subsection{Long-distance radiative coupling between the quantum dots}
We now try to give an answer to the question of how the radiative coupling depends on the inter-cavity distance in a PC dimer. The radiative coupling between the QDs is quantified by the tensor components of Eq. (\ref{Gtensor}), which are proportional to the Green's function evaluated at the QDs positions. Since the QDs are positioned in the center of each cavity, where the $E_y$ component is dominant, the dot-dot coupling is dominated only by the $G^{12}_{yy}$ component of the Green's tensor, which can be written for the two PC dimer modes as (see Eq. \ref{Gtensor}):

\begin{equation}\label{greeny}
G^{12}_{yy}(\omega)=\frac{2\pi\omega d^2}{\hbar}\left(\frac{E_{1,y}(\mathbf{r}_1)E^{*}_{1,y}(\mathbf{r}_2)}{\omega_1-i\frac{\gamma_1}{2}-\omega}+\frac{E_{2,y}(\mathbf{r}_1)E^{*}_{2,y}(\mathbf{r}_2)}{\omega_2-i\frac{\gamma_2}{2}-\omega}\right).
\end{equation}

We plot this effective coupling strength in Fig. \ref{fig:GreenF}(a) for the $30^{\circ}$ PC dimer as a function of frequency, for all interdot distances studied in the present work. As it is expected, when the QDs are in resonance with one of the PC dimer eigenfrequencies, the radiative coupling between the QDs is enhanced and it can remarkably reach values on the order of 1~meV or larger. However, the dot-dot coupling strength at these frequencies is surprisingly minimal at small inter-cavity distance, which are the cases where the coupling between the cavity modes is largest. As a counterintuitive consequence, the coupling strength between the QDs increases with the interdot distance after the smallest values of $d_c$, and remains relatively constant (small oscillations) at larger values of $d_c$. This can be easily verified in the Fig. \ref{fig:GreenF}(b), which shows the function $|G^{12}_{yy}(\omega)|$ at the values $\omega=\omega_1$, $\omega=\omega_2$, and $\omega=\omega_{av}=(\omega_1+\omega_2)/2$, 
respectively. The function $|G^{12}_{yy}(\omega)|$ evaluated at $\omega_{av}$ shows that even when the QDs are out of resonance from the dimer PC modes, the radiative dot-dot coupling is significant and increases with the interdot distance. For completeness, we show the respective calculations for the $60^{\circ}$ dimer in Figs. \ref{fig:GreenF}(c) and \ref{fig:GreenF}(d). The behavior of $|G^{12}_{yy}(\omega)|$ is very similar to the $30^{\circ}$ case, with the difference that the dot-dot coupling increases most rapidly with the interdot distance at small values of $d_c$, and due to the small splitting in the case of $d_c=4a$ we see a pronounced peak in $|G^{12}_{yy}(\omega)|$ at the mean frequency $\omega_{av}=(\omega_1+\omega_2)/2$. It is important to remind that these results are valid as long as the PC dimer mode splitting exceeds the photonic radiative linewidth, \textit{i.e.}, where the mode splitting can be spectrally resolved (strong cavity-cavity coupling condition). {Qualitatively similar results have been obtained for PC dimers in the $0^{\circ}$ and $90^{\circ}$ configurations, respectively (not shown here).}

These counterintuitive behaviors can be interpreted by analyzing the expression for the coupling constant $G^{12}_{yy}(\omega)$ in Eq. (\ref{greeny}). Since the fields are strongly localized in both cavity regions for all inter-cavity distances in the strong cavity-cavity coupling regime, the functions $E_y(\mathbf{r}_{\alpha})$ depend very weakly on the $d_c$ parameter; furthermore, the functions $E_{1,y}(\mathbf{r}_1)E^{*}_{1,y}(\mathbf{r}_2)$ and $E_{2,y}(\mathbf{r}_1)E^{*}_{2,y}(\mathbf{r}_2)$ are real due to the point symmetry of the structure with respect to the origin of coordinates, and have approximately the same value with opposite signs due to the opposite symmetries of the two modes [see Figs. \ref{fig:PC-structure}(e) and \ref{fig:PC-structure}(f)]. In this way, we can approximate the $|G^{12}_{yy}(\omega)|$ function as: 
\begin{equation}
|G^{12}_{yy}(\omega)|\approx\frac{2\pi d^2 |g^{12}|}{\hbar}\omega\left|\frac{\Delta_m-i\frac{\gamma_2-\gamma_1}{2}}{(\omega_1-i\frac{\gamma_1}{2}-\omega)(\omega_2-i\frac{\gamma_2}{2}-\omega)}\right|,
\end{equation}
where $|g^{12}|=|E_{1,y}(\mathbf{r}_1)E^{*}_{1,y}(\mathbf{r}_2)|\approx|E_{2,y}(\mathbf{r}_1)E^{*}_{2,y}(\mathbf{r}_2)|$, and $\Delta_{m}=\omega_2-\omega_1$. Neglecting the $\gamma$ terms when they do not contribute significantly to the sums, as well as their second order terms, we obtain the following trends for the effective coupling constant:
\begin{gather}\label{trendsG}
 |G^{12}_{yy}(\omega_1)|\propto Q_1,\\
 |G^{12}_{yy}(\omega_2)|\propto Q_2,\\
 |G^{12}_{yy}(\omega_{av})|\propto \frac{\omega_{av}}{\Delta_m},
\end{gather}
which qualitatively explain the results shown in Figs. \ref{fig:Q-factors} and \ref{fig:GreenF}. Eventually, at very large inter-cavity distance, where the mode splitting tends to zero and is much smaller than the mode radiative linewidth, the coupling constant $g^{12}$ tends to zero and consequently the dot-dot radiative coupling vanishes.

\begin{figure}[t]
  \begin{center}
    \includegraphics[width=0.45\textwidth]{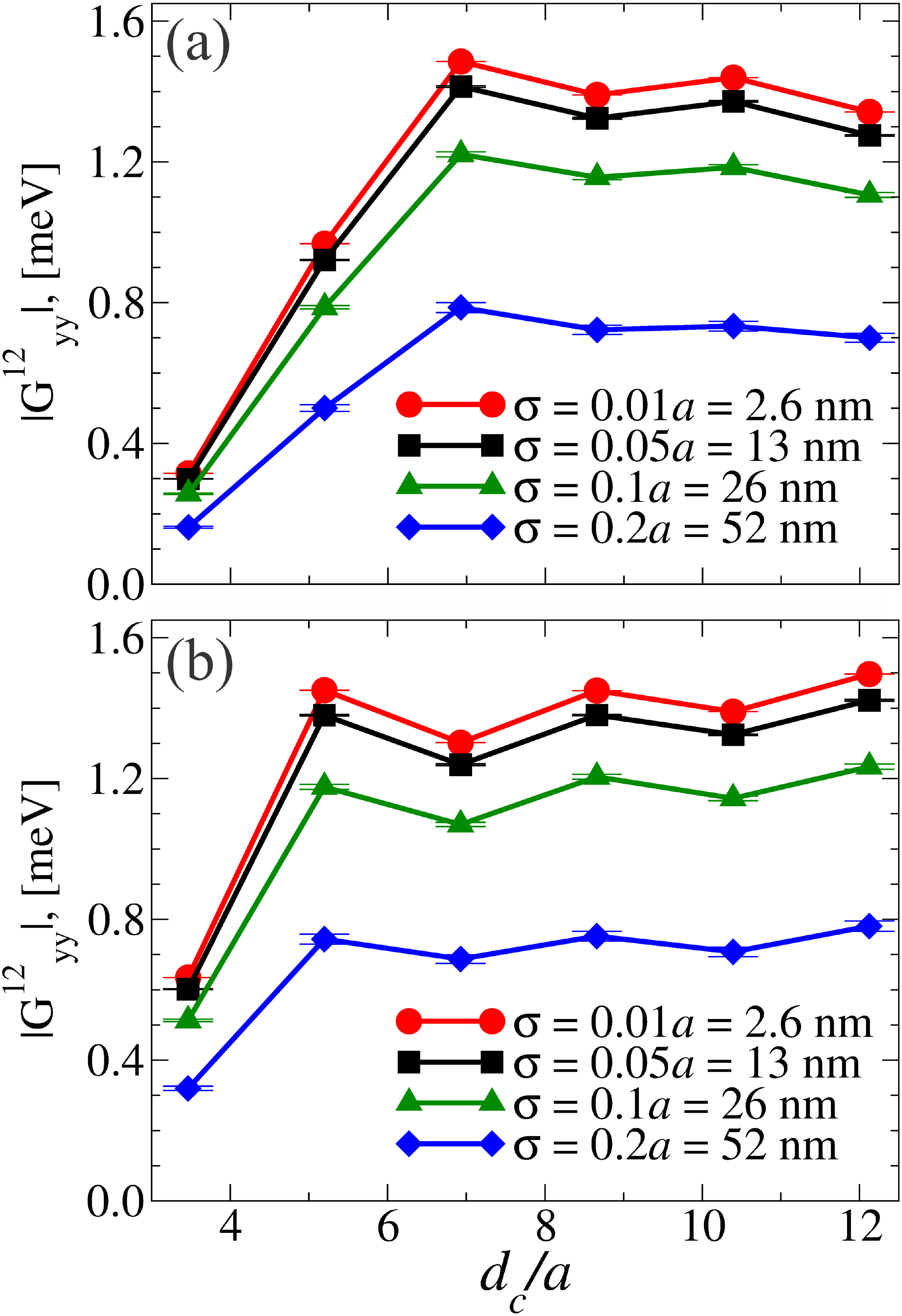}
  \end{center}
  \caption{
(Color online) (a) Absolute value of the effective inter-dot coupling strength at $\omega=\omega_1$, in the $30^{\circ}$ dimer, as a function of the cavity-cavity distance, for QDs positions generated randomly with a Gaussian probability distribution of variance $\sigma$. (b) The same as (a) with $\omega=\omega_2$. {The standard error is shown with vertical error bars}.}
  \label{fig:GreenF-statistics}
\end{figure}

\subsection{Non-ideal quantum dots positioning }
Coupling of the QDs with the PC dimer modes is maximized when the two QDs are exactly positioned at the centers of the two L3 cavities. The semiclassical formalism summarized in Sec. (\ref{model}) allows to study the effect of a non-ideal positioning of the QDs through a statistical analysis. To accomplish this, the position of one QD is generated by a two-dimensional Gaussian probability distribution with variance $\sigma$, and the corresponding position of the other QD is automatically determined since the dot-dot distance is maintained fixed. This model simulates in a realistic way possible misalignments in the writing stage of the photonic pattern with respect to the QDs, which can be deterministically positioned with a high degree of accuracy \cite{badolato2005,kevin07nat}. Figure \ref{fig:GreenF-statistics} shows the results of this analysis, where the absolute value of the Green's tensor component $G^{12}_{yy}(\omega)$ is studied as a function of the interdot distance for the $30^{\circ}$ dimer. 
Panels (a) and (b) correspond to the cases $|G^{12}_{yy}(\omega_1)|$ and $|G^{12}_{yy}(\omega_2)|$, respectively. The curves were averaged over 1000 realizations for each value of $\sigma$, {and the standard error is explicitly shown in the Figure with the corresponding error bars}. We see that the dot-dot coupling strength remains of the order of 1~meV for a statistical variance of $\sigma=$26 nm, which is comparable to the precision that can be achieved with modern sample-fabrication techniques \cite{reinhard2012nphot}. Furthermore, the coupling strength is sizable even for larger values of $\sigma$, as it can be seen from the figure. As an example, the coupling constant is of the order of  $0.7$~meV for $\sigma=52$~nm. We point out that in the present analysis the QDs are not necessarily positioned at the cavity centers, since their positions are generated randomly with a Gaussian distribution, and consequently the QDs will also couple to the $E_x$ component of the field. As a consequence, the other components of the Green's tensor might be 
non-negligible. Nevertheless, $|G^{12}_{yy}|$ is much larger and 
dominates for the values of $\sigma$ considered here. 
As a final remark, we notice that the results for the {$0^{\circ}$, $60^{\circ}$, and $90^{\circ}$ PC dimers} are equivalent to the $30^{\circ}$ case, and coupling strengths of the same order of magnitude can be obtained. 

\section{Conclusion}\label{conclusions}
We have studied the radiative coupling between two distant quantum dots embedded in the two cavities of photonic crystal molecules in planar waveguide geometry, or photonic crystal dimers, by using a semiclassical formalism based on the Green's tensor. The photonic eigenmodes are found by guided-mode expansion, which allows to estimate real and imaginary parts (losses) of the photonic eigenmodes, as well as the spatial mode profiles. Specifically, we have considered two L3 cavities made of three missing holes in a triangular lattice, in which field antinodes occur at each cavity center where the quantum dots can be placed. Parameters have been chosen to describe current systems typically fabricated with III-V semiconductors, such as InGaAs quantum dots in GaAs thin membranes. 

Irrespective of the coupling angle between the two cavities ($0^{\circ}$, $30^{\circ}$, $60^{\circ}$, $90^{\circ}$, respectively),  we have shown that the effective interdot coupling is enhanced when the QDs are in resonance with either of the two normal modes of the PC dimer. Under such resonance conditions, and in the strong cavity-cavity coupling regime, the interdot coupling strength is actually proportional to the quality factors of the normal modes (bonding or antibonding), and it can be of the order of 1~meV for the cases considered here, which is at least an order of magnitude larger than typical values achieved in one-dimensional systems \cite{minkov2013a}. Since these quality factors can also increase as a function of the inter-cavity distance, tending to the limiting value set by the isolated cavity mode (in the $10^5$ range for the present case), then the radiative coupling can also increase with distance. Moreover, since the quality factors remain approximately constant at large distances, the radiative coupling can also remain constant at inter-dot separation that is significantly larger than their characteristic emission wavelength. In addition, we have also shown that when the QDs are out of resonance from the PC dimer modes, the inter-dot radiative coupling is still significant and inversely proportional to the normal mode splitting between bonding/antibonding photonic states. Eventually, the mutual QDs coupling goes to zero when such normal mode splitting is blurred by their linewidth. Finally, with a statistical analysis of the non-ideal positioning of the QDs within their respective cavities, we could quantitatively conclude that with Gaussian variances comparable to the precision of modern sample fabrication techniques, the coupling strength is not strongly affected and remains of the order of 1~meV.

We notice that part of the conclusions drawn in the present work can be extended to any type of photonic dimer, in principle. In fact, while the mentioned 1 meV coupling strength is quantitatively valid for the specific systems considered here, it is still general the conclusion that the radiative coupling between quantum dots in photonic dimers remains constant even at significantly large inter-dot distances when two identical cavities are in the strong cavity-cavity coupling regime.
In fact, this is the main result of this work, which can be particularly relevant in view of possibly realizing PC dimers with normal mode splitting resolved even at very large distances \cite{noda_nphot2012}. The latter could definitely solve the issue of entangling distant qubits for applications in quantum information processing in an integrated photonic platform\cite{imamoglu99}, where independent optical manipulation of the two quantum emitters could be straightforward if their mutual distance significantly exceeds the resonant wavelength.

\section{Acknowledgements}
We acknowledge financial support from FAPEMIG, CAPES and CNPq, Brazil. This work was partially financed by PVE/CSF Project no. 407167/2013-7. 
The authors would like to thank L.~C. Andreani, A. Badolato, M. Gurioli, M. F. Santos, V. Savona for useful discussions.
J. P. Vasco would like to acknowledge the Photonics and Nanostructures group for the wonderful reception during his stay in Pavia.


\begin{thebibliography}{99}

%
% Intro and reviews
%
\bibitem{qp_review}
%For a recent review: "Quantum photonic technologies"
J. L. O'Brien, A. Furusawa, and J. Vu\v{c}kovi\'{c},
Nat. Photonics {\bf 3}, 687 (2009).

%
% Quantum dots as qubits
%
\bibitem{imamoglu99}
A. Imamo\v{g}lu, D. D. Awschalom, G. Burkard, D. P. DiVincenzo, D. Loss, M. Sherwin, and A. Small, 
% Quantum Information Processing Using Quantum Dot Spins and Cavity QED
Phys. Rev. Lett. {\bf 83}, 4204  (1999).

\bibitem{warburton_review}
For a recent review, see R. Warburton,
%Single spins in self-assembled quantum dots
Nat. Materials, {\bf 12}, 483 (2013). 

%
% Theoretical proposals for coupling distant QDs
%
\bibitem{parascandolo2005}
G. Parascandolo and V. Savona, 
% Long-range radiative interaction between semiconductor quantum dots
Phys. Rev. B {\bf 71}, 045335 (2005).

\bibitem{tarel2008}
G. Tarel, G. Parascandolo, and V. Savona, 
% Ultralong-range radiative excitation transfer between quantum dots in a planar microcavity
Phys. Status Solidi B {\bf 245}, 1085 (2008).

\bibitem{hughes2007prl}
S. Hughes, 
%Coupled-Cavity QED Using Planar Photonic Crystals 
Phys. Rev. Lett. {\bf 98}, 083603 (2007).

%
% Photonic crystal waveguides and resonators 
%
\bibitem{akahane2003}
Y. Akahane, T. Asano, B. S. Song, and S. Noda,
%``High-Q photonic nanocavity in a two-dimensional photonic crystal,''
{Nature} $\mathbf{425}$, 944 (2003).

\bibitem{notomi2004}
M. Notomi, A. Shinya, A. Mitsugi, S. Kuramochi, and H.Y. Ryu,
%``Waveguides, resonators and their coupled elements in photonic crystal slabs'',
Opt. Express {\bf 12}, 551 (2004). 

\bibitem{notomi_review}
For a review, see M. Notomi,
%``Manipulating light with strongly modulated photonic crystals.''
Rep. Prog. Phys. {\bf 73}, 096501 (2010).

%
% Quantum dots and cavity QED
%
\bibitem{andreani99prb}
L. C. Andreani, G. Panzarini, and J.-M. G\'{e}rard, 
% Strong-coupling regime for quantum boxes in pillar microcavities: Theory
Phys. Rev. B {\bf 60}, 13276 (1999).

%\bibitem{auffeves2010}
%A. Auff\`{e}ves, D. Gerace, J.-M. G\'erard, M. F. Santos, L. C. Andreani, and J.-P. Poizat. 
%%``Controlling the dynamics of a coupled atom-cavity system by pure dephasing.''
%Phys. Rev. B {\bf 81}, 245419 (2010).

\bibitem{yoshie2004}
T. Yoshie, A. Scherer, J. Hendrickson, G. Khitrova, H. M. Gibbs, G. Rupper, C. Ell, O. B. Shchekin, and D. G. Deppe, 
%Vacuum Rabi splitting with a single quantum dot in a photonic crystal nanocavity, 
Nature (London) {\bf 432}, 200 (2004).

\bibitem{badolato2005}
A. Badolato, K. Hennessy, M. Atat\"{u}re, J. Dreiser, E. Hu, P. M. Petroff, and A. Imamo\v{g}lu, 
% Deterministic Coupling of Single Quantum Dots to Single Nanocavity Modes
Science {\bf 308}, 1158 (2005).

\bibitem{kevin07nat}
K. Hennessy, A. Badolato, M. Winger, D. Gerace, M. Atat\"{u}re, S. Gulde, S. F\"{a}lt, E. Hu, and A. Imamo\v{g}lu,
% Quantum nature of a strongly coupled single quantum dot-cavity system
Nature (London) {\bf 445}, 896 (2007).

\bibitem{faraon08nphys}
A. Faraon, I. Fushman, D. Englund, N. Stoltz, P. Petroff, and J. Vu\v{c}kovi\'{c},
% Coherent generation of non-classical light on a chip via photon-induced tunnelling and blockade
Nat. Physics {\bf 4}, 859 (2008).

\bibitem{reinhard2012nphot}
A. Reinhard, T. Volz, M. Winger, A. Badolato, K. J. Hennessy, E. L. Hu, and A. Imamo\v{g}lu,
% Strongly correlated photons on a chip
Nat. Photonics {\bf 6}, 93 (2012).

%
% Photonic crystal caviies
%
\bibitem{hennessy2005}
K. Hennessy, A. Badolato, A. Tamboli, P. M. Petroff, E. Hu, M. Atat\"{u}re, J. Dreiser, and A. Imamo\v{g}lu, 
%Tuning photonic crystal nanocavity modes by wet chemical digital etching, 
Appl. Phys. Lett. {\bf 87}, 021108 (2005).

\bibitem{gerace05pnfa}
D.~Gerace and L.~C. Andreani,
%``Effects of disorder on propagation losses and cavity Q-factors in photonic crystal slabs,''
Photon. Nanostruct. Fundam. Appl. {\bf 3}, 120 (2005).

\bibitem{portalupi2011}
S. L. Portalupi, M. Galli, M. Belotti, L. C. Andreani, T. F. Krauss, and L. O'Faolain, 
% Deliberate versus intrinsic disorder in photonic crystal nanocavities investigated by resonant light scattering
Phys. Rev. B {\bf 84}, 045423 (2011).

\bibitem{tran2009}
N.-V.-Q. Tran, S. Combri\'e, and A. De Rossi,
%``Directive emission from high-Q photonic crystal cavities through band folding,''
Phys. Rev. B {\bf 79}, 041101R (2009).

%\bibitem{toishi2009}
%M. Toishi, D. Englund, A. Faraon, and J. Vu\v{c}kovi\'{c},
%%``High-brightness single photon source from a quantum dot in a directional emission nanocavity,''
%Opt. Express {\bf 17}, 14618 (2009).

\bibitem{portalupi2010opex}
S. L. Portalupi, M. Galli, C. Reardon, T. F. Krauss, L. O'Faolain, L. C. Andreani, and D. Gerace,
%`` Planar photonic crystal cavities with far-field optimization for high coupling efficiency and quality factor,''
Opt. Express {\bf 18}, 16064 (2010).

\bibitem{minkov2014scirep}
M. Minkov and V. Savona, 
% Automated optimization of photonic crystal slab cavities
Sci. Rep. {\bf 4}, 5124 (2014).

\bibitem{laiAPL2014}
 Y. Lai, S. Pirotta, G. Urbinati, D. Gerace, M. Minkov, V. Savona, A. Badolato, and M. Galli,
%``Genetically designed L3 photonic crystal nanocavities with measured quality factor exceeding one million,'' 
 {Applied Physics Letters}, {\bf 104}, 241101 (2014).
 
%
% distant QDs in phc circuits
%
\bibitem{yao2009}
P. Yao and S. Hughes,
%Macroscopic entanglement and violation of BellÕs inequalities between two spatially separated quantum dots in a planar photonic crystal system
Opt. Express {\bf 17}, 11505 (2009).

\bibitem{minkov2013a}
M. Minkov and V. Savona, 
% Radiative coupling of quantum dots in photonic crystal structures
Phys. Rev. B {\bf 87}, 125306 (2013).

\bibitem{minkov2013b}
M. Minkov and V. Savona, 
% Long-distance radiative excitation transfer between quantum dots in disordered photonic crystal waveguides
Phys. Rev. B {\bf 88}, 081303R (2013).

\bibitem{imamoglu2007}
A. Imamo\v{g}lu, S. F\"{a}lt, J. Dreiser, G. Fernandez, M. Atat\"{u}re, K. Hennessy, A. Badolato, and D. Gerace,
%Coupling quantum dot spins to a photonic crystal nanocavity
{J. Appl. Phys.}, {\bf 101}, 081602 (2007).

\bibitem{postigo2010}
E. Gallardo, L. J. Martinez, A. K. Nowak, D. Sarkar, H. P. van der Meulen, J. M. Calleja, C. Tejedor, I. Prieto, D. Granados, A. G. Taboada, J. M. Garcia, and P. A. Postigo,
% Optical coupling of two distant InAs/GaAs quantum dots by a photonic-crystal microcavity 
Phys. Rev. B {\bf 81}, 193301 (2010).

\bibitem{armando2008apl}
M. Benyoucef, S. Kiravittaya, Y. F. Mei, A. Rastelli, and O. G. Schmidt, 
% Strongly coupled semiconductor microcavities: A route to couple artificial atoms over micrometric distances, 
Phys. Rev. B {\bf 77}, 035108 (2008).

%
% Photonic crystal molecules
%
\bibitem{ishii2006}
S. Ishii, K. Nozaki, and T. Baba, 
%Photonic molecules in photonic crystals, 
Jpn. J. Appl. Phys. {\bf 45}, 6108 (2006).

%
% GME
%
\bibitem{andreani06prb}
L. C.~Andreani and D. Gerace,
%``Photonic crystal slabs with a triangular lattice of triangular holes investigated using a guided-mode expansion method,''
Phys. Rev. B {\bf 73}, 235114 (2006).

\bibitem{kee2003prb}
C. S. Kee, H. Lim, and J. Lee, 
%Coupling characteristics of localized photons in two-dimensional photonic crystals, 
Phys. Rev. B {\bf 67}, 073103 (2003).

\bibitem{caselli2012}
N. Caselli, F. Intonti, F. Riboli, A. Vinattieri, D. Gerace, L. Balet, L.H. Li, M. Francardi, A. Gerardino, A. Fiore, and M. Gurioli,
%%Antibonding ground state in photonic crystal molecule
Phys. Rev. B $\mathbf{86}$, 035133 (2012). 

\bibitem{atlasov2008}
K. A. Atlasov, K. F. Karlsson, A. Rudra, B. Dwir, and E. Kapon, 
%Wavelength and loss splitting in directly coupled photonic-crystal defect microcavities,
Opt. Express {\bf 16}, 16255 (2008).

\bibitem{chalcraftOpex}
A. R. A. Chalcraft, S. Lam, B. D. Jones, D. Szymanski, R. Oulton, A. C. T. Thijssen, M. S. Skolnick, D. M. Whittaker, T. F. Krauss, and A. M. Fox,
%Mode structure of coupled L3 photonic crystal cavities
Opt. Express {\bf 19}, 5670 (2011).

%
% Perturbation theory - GME
%
\bibitem{Sakoda}
T. Ochiai and K. Sakoda, Phys. Rev. B {\bf 64}, 045108 (2001).

%
% Far-field
%
%\bibitem{intonti2011prl}
%F. Intonti, F. Riboli, N. Caselli, M. Abbarchi, S. Vignolini, D. S. Wiersma, A. Vinattieri, D. Gerace, L. Balet, L. H. Li, 
%M. Francardi, A. Gerardino, A. Fiore, and M. Gurioli, 
%%YoungÕs Type Interference for Probing the Mode Symmetry in Photonic Structures, 
%Phys. Rev. Lett. {\bf 106}, 143901 (2011).
%
%\bibitem{brunstein2011apl}
%M. Brunstein, T. J. Karle, I. Sagnes, F. Raineri, J. Bloch, Y. Halioua, G. Beaudoin, L. Le Gratiet, J. A. Levenson, and A. M. Yacomotti, 
%%Radiation patterns from coupled photonic crystal nanocavities,
%Appl. Phys. Lett. {\bf 99}, 111101 (2011).

%
% Tuning
%
\bibitem{kicken2009opex}
H. H. J. E. Kicken, P. F. A. Alkemade, R. W. van der Heijden, F. Karouta, R. N\"{o}tzel, E. van der Drift, and H. W. M. Salemink, 
%Wavelength tuning of planar photonic crystals by local processing of individual holes, 
Opt. Express {\bf 17}, 22005 (2009).

\bibitem{caselli2012apl}
N. Caselli, F. Intonti, C. Bianchi, F. Riboli, S. Vignolini, L. Balet, L. H. Li, M. Francardi, 
A. Gerardino, A. Fiore, and M. Gurioli, 
%Post-fabrication control of evanescent tunnelling in photonic crystal molecules, 
Appl. Phys. Lett. {\bf 101}, 211108 (2012).

\bibitem{intonti2012apl}
F. Intonti, N. Caselli, S. Vignolini, F. Riboli, S. Kumar, A. Rastelli, O. G. Schmidt, M. Francardi, A. Gerardino, L. Balet, 
L. H. Li, A. Fiore, and M. Gurioli, 
%Mode tuning of photonic crystal nanocavities by photoinduced nonthermal oxidation, 
Appl. Phys. Lett. {\bf 100}, 033116 (2012).

\bibitem{waks2013apl}
T. Cai, R. Bose, G. S. Solomon, and E. Waks,
%Controlled coupling of photonic crystal cavities using photochromic tuning
Appl. Phys. Lett. $\mathbf{102}$, 141118 (2013). 

\bibitem{caselli2014opex}
N. Caselli, F. Intonti, F. Riboli, and M. Gurioli, 
%Engineering the mode parity of the ground state in photonic crystal molecules,Ó 
Opt. Express {\bf 22}, 4953 (2014).

\bibitem{yacomotti2014}
S. Haddadi, P. Hamel, G. Beaudoin, I. Sagnes, C. Sauvan, P. Lalanne, J. A. Levenson, and A. M. Yacomotti,
%Photonic molecules: tailoring the coupling strength and sign,
Opt. Express {\bf 22}, 12359 (2014).

\bibitem{noda_nphot2012}
Y. Sato, Y. Tanaka, J. Upham, Y. Takahashi, T. Asano, and S. Noda,
% Strong coupling between distant photonic nanocavities and its dynamic control
Nat. Photonics {\bf 6}, 56 (2012).

%
% Current interest for PC molecules
%
\bibitem{gerace_josephson}
D. Gerace, H. E. T\"{u}reci, A. Imamo\v{g}lu, V. Giovannetti, and R. Fazio,
% The quantum optical Josephson interferometer
Nat. Physics {\bf 5}, 281 (2009).

\bibitem{ferretti2010}
S. Ferretti, L.C. Andreani, H.E. T\"{u}reci, and D. Gerace,
% Photon correlations in a two-site nonlinear cavity system under coherent drive and dissipation
Phys. Rev. A {\bf 82}, 013841 (2010).

\bibitem{savona10prl}
T. C. H. Liew and V. Savona, 
% Single photons from coupled modes
Phys. Rev. Lett. {\bf 104}, 183601 (2010).

\bibitem{bamba}
M. Bamba, A. Imamo\v{g}lu, I. Carusotto, and C. Ciuti,
%``Origin of strong photon antibunching in weakly nonlinear photonic molecules.''
Phys. Rev. A {\bf 83}, 021802(R) (2011).

\bibitem{arka2012prb}
A. Majumdar, A. Rundquist, M. Bajcsy, and J. Vu\v{c}kovi\'{c},
%Cavity quantum electrodynamics with a single quantum dot coupled to a photonic molecule
Phys. Rev. B {\bf 86}, 045315 (2012).

\bibitem{hakan_arxiv}
C. Aron, M. Kulkarni, and H. E. T\"{u}reci,
%Steady-state entanglement of spatially separated qubits via quantum bath engineering
arXiv:1403.6474 (2014).

\bibitem{gerace04pnfa}
L.~C. Andreani, D.~Gerace and M.~Agio,
% Gap maps, diffraction losses, and exciton–polaritons in photonic crystal slabs
Photon. Nanostruct. Fundam. Appl. {\bf 2}, 103 (2004).


%\bibitem{portalupi2011}
%S. L. Portalupi, M. Galli, M. Belotti, L. C. Andreani, T. F. Krauss, and L. O'Faolain, 
%Phys. Rev. B {\bf 84}, 045423 (2011).

%\bibitem{derossi2008}
%S. Combri\'e, A. De Rossi, N.-Q.-V. Tran, and H. Benisty,
%%GaAs photonic crystal cavity with ultrahigh Q: microwatt
%%nonlinearity at 1.55 $\mu$m.
%Opt. Lett. $\mathbf{33}$, 1908 (2008).


\bibitem{hopfield1958}
J. J. Hopfield, 
% Theory of the Contribution of Excitons to the Complex Dielectric Constant of Crystals
Phys. Rev. {\bf 112}, 1555 (1958).



\end{thebibliography}
\end{document}